\documentclass[aps,prl,notitlepage,superscriptaddress,reprint]{revtex4-2}
\usepackage{ucs}
\usepackage[utf8x]{inputenc}
\usepackage{amsmath}
\usepackage{amsfonts}
\usepackage{amssymb}
\usepackage{amsthm}
\usepackage{fontenc}
\usepackage[dvips,pdftex]{graphicx}
\usepackage{setspace}
\usepackage{color}
\usepackage[dvipsnames]{xcolor}
\usepackage{graphicx}

\usepackage{hyperref}

\usepackage{multirow}
\usepackage{array}
\usepackage{caption}
\usepackage{subcaption}
\usepackage{tikz}
\usepackage{multimedia}
\usepackage{float}
\usepackage{rotating}
\usepackage[outdir=./]{epstopdf}

\makeatletter
\@addtoreset{chapter}{part}
\makeatother

\newcommand{\dt}{\,\text{d}}

\newcommand{\pd}{\partial}

\newcommand{\lpow}[2]{\,{}^{#1}\!#2}

\newcommand{\ptwo}{\lpow{2}{P}}
\newcommand{\Qtwo}{Q}
\newcommand{\varpitwo}{\lpow{2}{\tilde{\varpi}}}

\DeclareGraphicsExtensions{.pdf,.eps,.jpg,.png}

\hypersetup{linktocpage}

\begin{document}
\begin{abstract}
We derive a complete, coarse grained, continuum model of the 2D vertex model. It is applicable for different underlying geometries, and allows for analytical analysis of an otherwise numerical model. Using a geometric approach and out--of--equilibrium statistical mechanics, we calculate both mechanical and dynamical instabilities within a tissue, and their dependence on different variables, including activity, and disorder. Most notably, the tissue's response depends on the existence of mechanical residual stresses on a cellular level. Thus, even freely growing tissues may exhibit a growth instability depending on food consumption. Using this geometric model we can readily distinct between elasticity and plasticity in a growing, flowing, tissue.
\end{abstract} 

\title{Instabilities and geometry of growing tissues}
\author{Doron \surname{Grossman}}
\email[]{doron.grossman@college-de-france.fr}
\affiliation{College de France}
\author{Jean-Francois \surname{Joanny}}
\affiliation {College de France}

\date{\today}
\maketitle

Cellular media {such as tissues are often described using the so-called vertex model, in which cells are described as confluent polygons or polyhedra. In this paper, we consider 2-dimensional tissues, which are a good representation of squamous epithelia. Each cell is assigned a preferred  area and perimeter as sketched in Fig~. \ref{fig: localized}.a  	\cite{honda1983geometrical,fletcher2014vertex}. The energy 	of the tissue depends on the 	difference of the actual area and perimeter from their preferred values.
\begin{align}\label{eq: vertex model 1}
	E & =  \sum_i  \hat{k}_A \left(A_i - \hat{A}^0_i\right)^2 + k_P \left(\ptwo_i -\ptwo^0_i\right)^2.
\end{align}
We use a slightly modified model replacing 
the actual perimeter by 	$\ptwo = \sum_e \ell_e^2$, where $\ell_e$ is the length of edge $e$ of the cell. This model agrees with the standard formulation of the vertex model to linear order, and we expect it to exhibit a qualitatively similar behaviour beyond linearity, while it significantly simplifies further calculations (see appendix A). The area and perimeter moduli are positive properties of the cells.

In a tissue, cells divide, die, and move past each other exchanging neighbors via $T_1$ transition, as shown in fig. \ref{fig: localized}). All these transformations change the edge-network topology, and can relax internal stresses. Recent studies show subtle solid-solid transitions, due to the appearance of soft-deformation modes, depending both on the ratio between the reference perimeter and area $\ptwo^0/A^0$, and the magnitude of disorder. \cite{farhadifar2007influence,staple2010mechanics,moshe2018geometric,sahu2019nonlinear}.The discrete vertex model typically requires numerical studies \cite{chiou2012mechanical,merkel2017using,popovic2021inferring,guirao2015unified}. Other theoretical descriptions of tissues include phenomenological continuum models  \cite{ranft2010fluidization,hannezo2014theory}, or otherwise neglecting relaxation and plasticity \cite{murisic2015discrete,moshe2018geometric}. One of the most significant difficulties is the distinction between solid-like (elastic) behavior, and fluid-like behavior \cite{guirao2015unified,merkel2017triangles}. 
Here we present a rigorous derivation of a continuum vertex model using an intrinsic, geometric, approach. The result is an easily generalized, 2-dimensional model, that takes into account geometry and out - of - equilibrium statistical mechanics, which exemplifies the difference between elasticity and the plasticity in the vertex model, and allows for an analytical treatment of the various topological processes allowed in the tissue.


Given a configuration $\vec{f}\left(x^\mu\right)$, describing the position of cells vertices at coordinates $x^\mu$, the (induced) metric is defined by $g_{\mu\nu} = \pd_\mu \vec{f} \cdot \pd_\nu \vec{f}$. 
Assuming cells' size is the smallest scale in the system, we approximate
the distance between two neighboring vertices by $\ell_{12}^2 \simeq g_{\mu\nu}\Delta x_{12}^\mu\Delta x_{12}^\nu$  ($\Delta x_{12}^\mu = x_2^\mu - x_1^\nu$ being the coordinate difference between vertices). The perimeter is $\ptwo_i= \sum_e \ell_e^2 = g_{\mu\nu}(x_i) Q^{\mu\nu}_i$. We define the network tensor:
\begin{align}\label{eq: network tensor}
	Q_i^{\mu\nu} = \sum_e \Delta x_e^{\mu} \Delta x_e^\nu,
\end{align}
where the sum is taken over all edges $e$ of the $i^{th}$ cell.

The cell area itself is not a simple function of $Q^{\mu\nu}_i$, still it can be written as $A_i = s_i \sqrt{g(x_i)} \sqrt{Q_i}$ where $g =\det g_{\mu\nu}$, $Q = \det Q^{\mu\nu}$, and $s_i(Q^{\mu\nu}) = \frac{A_i}{\sqrt{g(x_i)Q_i}} \neq 0$, is a finite correction term that depends smoothly on $Q^{\mu\nu}$. This factor can be factored out, renormalizing the area term in a controlled manner. The energy may then be written as
\begin{align}\label{eq: energ metric}
	E & = \sum_i {k}_A\left(\sqrt{g Q_i} -{A}^0_i\right)^2 + k_P\left(g_{\mu\nu}Q^{\mu\nu}_i -\ptwo^0\right)^2,
\end{align}
where ${k}_A$ and ${A}^0_i$ are the re-normalized values.

	In a continuum model \cite{gorban2006basic}, a tissue is described by a density function $\Psi\left(x,A^0,\ptwo^0,Q^{\mu\nu}\right)$, normalized so that $\int \Psi \dt[Q,A^0,\ptwo^0] =\sqrt{g} \rho$, where $\rho$  is the cell density and $\dt[Q,A^0,\ptwo^0] = \dt Q^{\mu\nu} \dt A^0 \dt \ptwo^0$ is the integration measure over the dynamic fields. The energy of the continuum model reads:
\begin{align}\label{eq: continuum vertx model}
	E & =  \int  \int  \Psi  \left\{ k_A \left(\sqrt{g Q} -A^0 \right)^2  \right. \\ \nonumber & \left. + k_P \left(g_{\mu\nu}Q^{\mu\nu}- \ptwo^0\right)^2 \right\} \dt[Q,A^0,\ptwo^0] \dt^2 x 
\end{align}
 where $\dt^2 x = \dt x^1\dt x^2$ is the integration measure over coordinates. Completing the integration over the dynamic fields, we get 
\begin{align}\label{eq: avg energy}
	E & =  \int \rho(x) \left\{k_A \left( A\left( x \right) - A^0\left( x \right) \right)^2   \right. \\ \nonumber & \left. +k_P \left(\ptwo\left( x \right) - \ptwo^0\left( x \right) \right)^2  +
	k_A \left( \Delta A^2 + \Delta {A^0}^2  \right)  \right. \\ \nonumber & \left. +  k_P\left(\Delta\ptwo^2 + \Delta{\ptwo^0}^2\right) \right\}\sqrt{g} \dt x 
\end{align}
where the field $f(x)= \langle f \rangle = \frac{\int \Psi f \dt[Q,A^0,\ptwo^0] }{\int \Psi \dt[Q,A^0,\ptwo^0] }$ is the average of the variable $f$, and $\Delta f^2 = \langle f^2 \rangle -\langle f \rangle ^2$ is the variance, which include spatial derivative as well statistical, local, variance (see appendix B). 

We assume that the elastic stress $\sigma_{el}^{\mu\nu} = -\frac{\delta E}{\delta g_{\mu\nu}}$ is a fast relaxing variable, and is divergence free (in the absence of external forces),
\begin{align}\label{eq: elastic equation}
		\nabla_\mu \sigma^{\mu \nu}_{el} = 0.
\end{align}
However, the tissue may also grow and relax internal stress. This is achieved via two type of processes. Solid-like relaxation only involves non-topological changes of the cellular network, such as growth (change in reference area), and elongation (change in reference perimeter term). Fluid-like relaxation, involves topological changes and induces cellular flow and diffusion. We consider  proliferation
(division and apoptosis), and $T_1$ transitions.

When a tissue behaves as a solid, a coordinate system is always well defined, and tissue deformations correspond directly to changes in the metric $g_{\mu\nu}$. These may be caused either by external forces, $F^\mu_{ext}$, so that $\nabla_\nu \sigma^{\mu \nu}_{el}=F^\mu_{ext}$,  or by internal changes in the reference values, which can be calculated
by solving \eqref{eq: elastic equation}, as $\sigma^{\mu \nu}_{el}$ evolves in time with $A^0$ and $\ptwo^0$.

As cells flow, i.e. - change their relative positions, one may need to redefine the coordinate system of the tissue. In the absence of any external reference, this may seem as a non-trivial feat. Yet, if topological changes occur in a small region, while neighboring tissue does not, one may keep the coordinates on the non-changing surrounding tissue, and deduce that the coordinates in the region of transition have remained the same as well  (see fig. \ref{fig: localized} (b)-(d)). Rather, the quantity that has changed is the network tensor, $Q^{\mu\nu}(x)$, at that point. This means that eq. \eqref{eq: elastic equation} is still valid and the metric is still well defined, even when $\sigma_{el}^{\mu\nu}$ non - trivially changes via its dependence on the network tensor $Q^{\mu\nu}$. Note, that within this Lagrangian view, the average cellular flow, as measured by a Lagrangian (co - moving) viewer is zero. To put simply, this happens since the mean flow involves large configuration deformations which may be regarded as change in the metric $g_{\mu\nu}$. Flow is thus described by change in the energy minimizing metric (which solves eq. \eqref{eq: elastic equation}) 
\begin{figure}
	\centering
	\includegraphics[width=0.85\columnwidth]{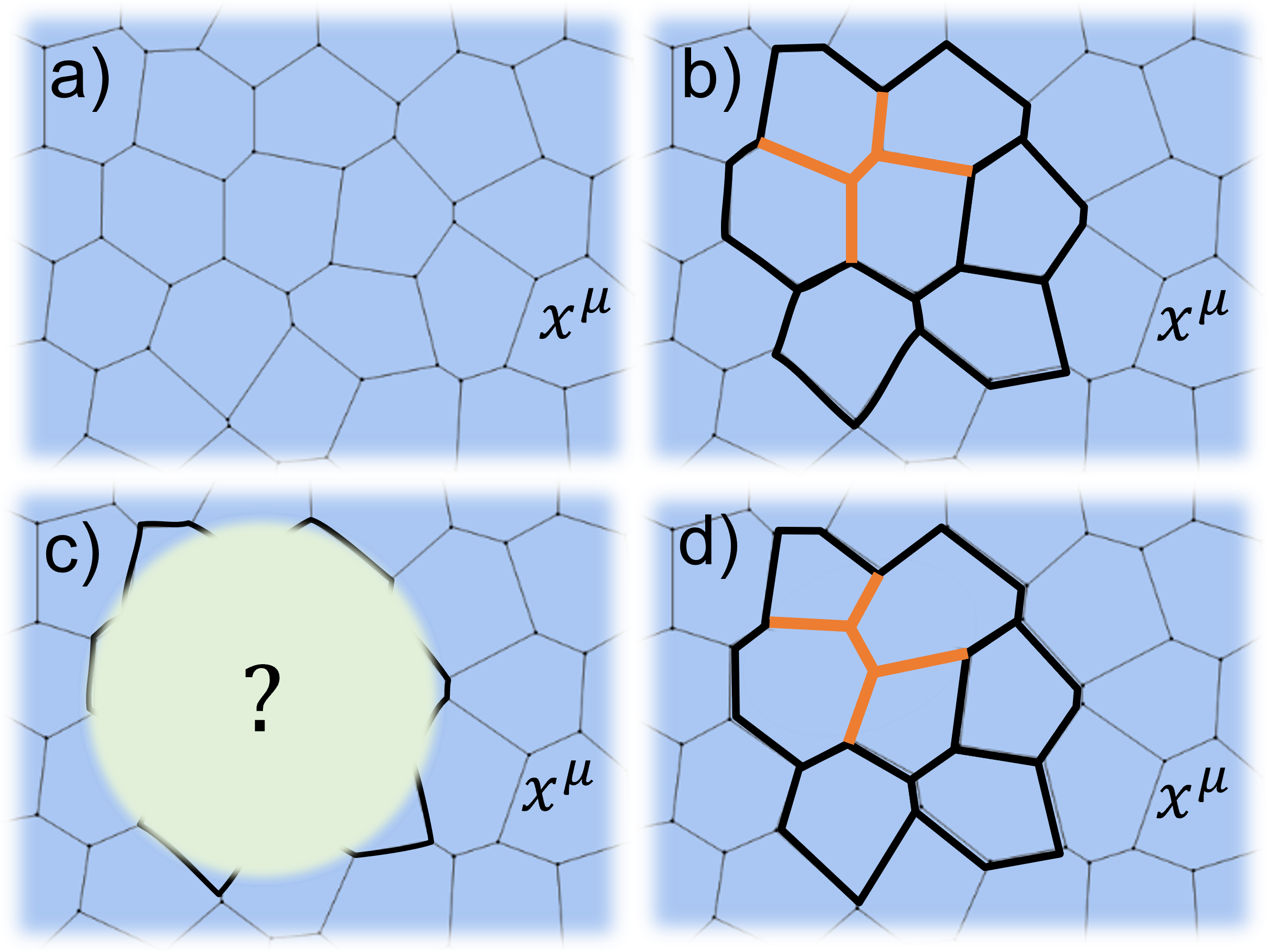}
	\caption{\label{fig: localized}  (a) A cellular tissue before any transition happens, it is easy to assign coordinates and metric. (b)-(d) A topoligical transition (in this case a $T_1$), outside obscured region the system remains a solid. (d) The final netwrok.}
\end{figure}

Tissue fluid-dynamics are described using a Fokker - Plank equation for the coarse grained density function $\Psi$, over coordinates and variables (a complete expression, including solid processes, is given in appendixes B, C):
\begin{align}\label{eq: Fokker Planck Final}
	\pd_t \Psi =&   \sum_{n} \left\{  \nabla_\mu\nabla_\nu \left[D_{n,x}^{\mu\nu} \Psi\right] -\frac{\pd}{\pd \Qtwo^{\mu\nu}}\left[\pd_t \Qtwo^{\mu\nu}_n \Psi\right] \right. \\ \nonumber
	&  + \left. \frac{\pd^2}{\pd \Qtwo^{\mu \nu}\pd \Qtwo^{\lambda \sigma}} \left[D_{i,\Qtwo}^{\mu\nu\lambda\sigma}\Psi\right]\right\}. \\ \nonumber
\end{align}
The first term is the usual Fokker - Plank diffusion term with vanishing mean flow, $D_{n,x}^{\mu\nu}$ is the diffusion coefficient stemming from the n$^{th}$ type topological process. The second term accounts for the change in network topology, the  third term is noise. The sum is taken over division ($d$), apoptosis ($a$), and $T_1$ transitions.

Eq. \eqref{eq: Fokker Planck Final} must be completed by the requirement that the local cellular density changes when cells divide or die: $\pd_t \rho =  \int  \pd_t \Psi \dt[A^0, \ptwo^0, \{\ell\}]  + \gamma\rho - \frac{\rho}{2}g^{\mu\nu} \pd_t g_{\mu\nu}+\zeta_\rho$, where the proliferation rate is $\gamma = \frac{1}{2} Q^{-1}_{\mu\nu} \left(\pd_t Q^{\mu\nu}_d + \pd_t Q^{\mu\nu}_a \right)$. We
consider a white noise with statistics $\langle \zeta_\rho (t,x) \zeta_\rho (t',x')\rangle  = D_\gamma \rho \delta(t-t')\delta(x-x')$ directly related to proliferation noise $D_{n,Q}^{\mu\nu\alpha\beta}$. 

Using eq. \eqref{eq: Fokker Planck Final} one can write the energy generation rate
\begin{align}\label{eq: Enery_prod_simp}
	\frac{dE}{dt} = & \int  \left\{\mu \gamma  +\eta_{\mu\nu} \pd_t \Qtwo^{\mu\nu} - \sigma_{el}^{\mu\nu}\pd_t g_{\mu\nu}  + \eta_D \right\} \rho\sqrt{g} \dt x,
\end{align}
where $\mu$ may be interpreted as a cell chemical potential,
$\sigma_{el}^{\mu\nu}$ is the elastic stress, $\eta_{\mu\nu}$ is the dissipation 
associated with the change in network, and $\eta_D$ is the a dissipation term due to 
diffusion (see appendix C for their full expressions). Note again that the 
proliferation rate $\gamma$ is directly related to $\pd_t Q^{\mu\nu}$ and is not 
an independent variable.  
Using an Onsager approach, we may now relate the unknown flux ($\pd_t 
Q^{\mu\nu}$) to the thermodynamic forces $\mu,\sigma_{el},\eta_{\mu\nu}$ 
which have an explicit expression using  in terms of our variables.  

In the following, we make two important simplifications. First, we perform a mean field approximation considering the local average values of all fields and we ignore terms involving the spatial derivatives of the fields. Also, for simplicity we assume that the fields are uniform, limiting ourselves to planar configurations. Second, we use a minimal relaxation model in which the Onsager coefficient matrix is diagonal, with the exception of the active contributions, which are added to all fluxes. A more complex treatment is left for future work.



The mean energy per cell reads 
\begin{align}\label{eq: mean energy disorder}
     E/N_{cell} = & k_A \left(\sqrt{g Q}-A^0\right)^2 + k_P \left(g_{\mu\nu}Q^{\mu\nu} - \ptwo^2 \right)^2 \\ \nonumber
     				&  + E_{\mu\nu\alpha\beta} \left\langle \Delta Q^{\mu\nu} \Delta Q^{\alpha\beta} \right\rangle,
 \end{align}
where $N_{cell}= \frac{\int \sqrt{g} \dt S}{A_{cell}}$, $\Delta E_{\mu\nu\alpha\beta}=k_A \sqrt{g \Qtwo} \Huge[ \Large(\sqrt{g 
\Qtwo}- \frac{1}{4}\Large) \Large(\Qtwo^{-1}_{\mu\nu}\Qtwo^{-1}_{\alpha\beta} - 
\Qtwo^{-1}_{\mu\alpha}\Qtwo^{-1}_{\nu\beta}\Large) + \frac{1}{4} \Qtwo^{-1}
_{\mu\alpha}\Qtwo^{-1}_{\nu\beta} \Huge] + k_P g_{\mu\nu}g_{\alpha \beta} $ 
is the expansion ofthe energy to second order in the fluctuations of  
$Q^{\mu\nu}$ with respect to its average value.  Here $N_{cell} \propto \frac{1}{\sqrt{Q}}$ is the total number of cells in the tissue at a given time.

In our  minimal scheme, Onsager relations are obtained for the network 
tensor, as the metric tensor is considered a fast variable. The dynamics of the network tensor are composed of a relaxation term and an active term: 
\begin{align}\label{eq: Q rule}
	\pd_t Q^{\mu\nu} = - \Gamma^{\mu\nu\alpha\beta} \eta_{\alpha \beta} - \xi Q^{\mu\nu}.
\end{align}
The first term is relaxation, with  Onsager coefficient $\Gamma^{\mu\nu\alpha\beta}$ and the conjugate force to the rate of change of the network tensor
$\eta_{\alpha \beta} = \frac{1}{2\sqrt{Q}}\left [k_A \left(g Q -{A^0}^2 \right) - k_P 
\left(g_{\mu\nu}Q^{\mu\nu}-\ptwo^0\right)^2 \right]Q^{-1}_{\alpha \beta} + 
\frac{2}{\sqrt{Q}}k_P \left(g_{\mu\nu}Q^{\mu\nu} - \ptwo^0\right)g_{\alpha\beta}
$. The second active term describes a constant proliferation rate 
depending on external energy supply, with $\xi>0$ for a growing tissue. A 
homeostatic state, is reached whenever $\pd_tQ =0$, as this indicates that 
the network, on average, does not change. The Onsager tensor, $
\Gamma^{\mu\nu\alpha\beta}$ may depend strongly on the topological process involved  (appendix D). The simplest tensorial form allowed by symmetry is
\begin{align}\label{eq: upsilon}
	\Gamma^{\mu\nu\alpha\beta} = H_1 \left(\frac{1}{2}\left(Q^{\mu\alpha}Q^{\nu\beta}+Q^{\nu\alpha}Q^{\mu\beta}\right) + H Q^{\alpha \beta} Q^{\mu\nu}\right) ,
\end{align}
where the rate constant $H_1$ is taken identical for all the 
topological processes (and can be set to $\upsilon=1$). The dimensionless 
parameter $H \geq -\frac{1}{2}$ is 
process dependent. For a pure $T_1$ transition, in which the cell area does 
not change on average,  $H = -\frac{1}{2}$, any $H> -\frac{1}{2}$ 
corresponds to 
some proliferation. 
While other forms of $\Gamma^{\mu\nu\alpha\beta}$ are allowed,  it is only 
this form with $H=- \frac{1}{2}$ that allows for "pure shear" where 
the area does 
not change, in any geometry. It is therefore very useful 
to use this same form 
in other cases as well. Additionally, it allows for a simple 
"one constant" (or 
average) limit by considering a single $\Gamma^{\mu\nu\alpha\beta}$ 
describing all processes at once.

\paragraph{Mechanics-} We first study a free, solid, tissue, 
for which $Q^{\mu\nu}$ does not change. Any deformation is then 
due to a change of the metric $g_{\mu\nu}$. The elastic stress in the tissue 
vanishes, so that $\sigma_{el}^{\mu\nu}= 0$ 
is an energy minimizing solution. Since both $Q^{\mu\nu}$ 
and $g_{\mu\nu}$ are positive definite tensors, one can work in a coordinate system in which $Q^{\mu
\nu} = \delta ^{\mu\nu}$ . Additionally, without loss of generality, 
one can choose the energy scale so that $k_A =1$ and the length scale 
so that $A^0=1$. Solving for $g_{\mu\nu}$, yields, the eigenvalues of 
$g_{\mu\nu}$, $g_1$ and $g_2$:
\begin{align}\label{eq: metric mech}
	\left.\begin{array}{cc}
	g_1 =\frac{1}{g_2} = \frac{ \ptwo^0  \pm \sqrt{ {\ptwo^0 }^2 -4}}{2}, & 
	\ptwo^0 \geq 2,  \\ \
	g_1 ={g_2} = \frac{1+2 k_P \ptwo^0}{1+4 k_P}, & \ptwo^0 < 2
	\end{array} \right.
\end{align}
These are the expected results known \cite{staple2010mechanics,moshe2018geometric,popovic2021inferring} for soft -- solids (in the limit $
\ptwo > 2$), where cells assume an elongated shape, along a 
spontaneously chosen direction, and for hard -- solids when $
\ptwo <2$) where cells are isotropic, and are under internal residual 
stress as easily seen from the fact that  $E\neq0$. The case $\ptwo=2$ corresponds to a marginal solid.
 
\paragraph{Fluctuating mechanics-} The effects of fluctuations 
depend, to some extent, on the nature of the disorder. We 
considered two archetypical examples, assuming that the 
correlation between different components of $Q^{\mu\nu}$ vanish.  
The first one is isotropic fluctuations, $\langle \Delta Q^{\mu\nu}\Delta 
Q^{\alpha \beta} \rangle =  \frac{\phi}{2}  \left(g^{\mu\alpha}g^{\nu \beta}
+g^{\nu\alpha}g^{\mu \beta}\right)$, $\phi$ being the fluctuation 
amplitude. This is a very intuitive model, suggesting that disorder is 
related to tissue shape. The second example assumes that whatever 
mechanisms govern the network $Q^{\mu\nu}$ gives rise to fluctuations, 
and thus the fluctuation should be should be proportional to $ \frac{\phi}
{2}  \left(O^{\mu\alpha}O^{\nu \beta}+O^{\nu\alpha}O^{\mu \beta}\right)$, 
where $\left(O^2\right)^{\mu\nu} = Q^{\mu\nu}$. We term this second 
example "multiplicative".  In both cases, the resulting effect is a 
shift of the critical transition between an isotropic and symmetry-broken 
tissues to a value $\ptwo^0_{crit} >2$. 
\begin{align}\label{eq: P0 crit}
	\ptwo^0_{crit} \simeq  2+ \phi \times \left\{ \begin{array}{cc}
		\frac{1+6 k_P}{2 k_P}  & \text{isotropic} \\
		\frac{1+16 k_P}{8 k_P} & \text{multiplicative}\\
	\end{array} \right. 
\end{align}
These results are in accordance with other similar accounts of changing the shape parameters transition between ordered and disordered tissues \cite{bi2015density,sahu2019nonlinear}, suggesting that fluctuations play a significant role in tissue mechanics and dynamics.

\paragraph{External force-} Within the mean field approach, one can calculate the non-linear responses to an external force, easily recovering the results of Ref. \cite{moshe2018geometric} (see appendix E), and calculate further non-linear response coefficients such as Poisson's ratio.

\paragraph{Relaxation Dynamics-}
The model also allows for an analytical calculation of  tissue dynamics. We first consider a tissue strongly adhered to a solid 
substrate, in the absence of growth $\xi =0 $. As the actual size of the tissue is assumed given, we choose a constant metric, 
$g_{\mu\nu} = \delta_{\mu\nu}$, and let $Q^{\mu\nu}$ evolve. In this case, internal elastic stresses may develop, but are balanced by forces from the substrate.  Working with the eigenvalues of $Q^{\mu\nu}$, $q_1$ and $q_2$, and 
using eq. \eqref{eq: upsilon}

\begin{align}\label{eq: Q eq}
	\dot{q}_i = & -\frac{\left(1 + 2 H\right) q_i  }{2 \sqrt{q_1 q_2}}\left[q_1 q_2 -1- k_P \left(q_1+q_2 -\ptwo^0\right)^2\right]  \\ \nonumber &
	- \frac{2}{\sqrt{q_1 q_2}} k_p \left(q_1+q_2-P^0\right)\left[q_i + H \left(q_1 + q_2\right) \right] q_i,
\end{align}
where $i\in \left(1,2\right)$. Stable solutions depend on $\ptwo^0$ (and therefore on the existence of residual stresses)

\begin{align}
	\left. \begin{array}{cc}
		q_1=1/q_2 = \frac{ \ptwo^0  \pm \sqrt{ {\ptwo^0 }^2 -4}}{2}, &\ptwo^0 \geq 2 \\
		q_1 =q_2 =\sqrt{\frac{1+k_p {\ptwo^0}^2}{1+ 4 k_P}}, & \ptwo^0 <2
	\end{array} \right.
\end{align}
Interestingly, the solution for $\ptwo^0 >2$ corresponds to the same 
solution (in different coordinates) as the pure mechanical problem. The 
solution for $\ptwo^0  <2$ differs. The case of a tissue with no proliferation is singular because the only topological transitions are the pure $T_1$ transitions, which conserve area. It is treated in appendix F

\paragraph{Growing tissue-} A tissue grows if the active growth rate $\xi >0$.  The number of cells is $N_{cell} \propto \frac{1}{\sqrt{Q}}$ so that  $\pd_t N_{cells} = -\frac{N_{cells}}{2} Q^{-1}_{\mu\nu}\pd_tQ^{\mu\nu}$.  When considering a free tissue, one has to find both the metric $g_{\mu\nu}$ and the network tensor $Q^{\mu\nu}$, by solving equations \eqref{eq: elastic equation} and \eqref{eq: Q rule} simultaneously.
For a free tissue we get $\pd_t N_{cells} = \gamma N_{cells}$. The growth rate $\gamma$ depends on the existence of residual stresses in the tissue:
\begin{align}
	\gamma = \left\{ \begin{array}{cc}
	 \xi, & \ptwo^0 \geq 2 \\
	\xi -\frac{1+2H}{2\sqrt{Q}}\mathcal{E}, & \ptwo^0 <2
\end{array} \right.
\end{align}
where $\mathcal{E}= \left(\frac{1+2k_P \ptwo^0}{1+4k_P}-1\right)^2 + k_P \left(2\frac{1+2k_P \ptwo^0}{1+4k_P} - \ptwo^0\right)^2 $ is the average residual energy per cell. The active growth rate, $\xi$ must be large enough to allow continuing tissue growth, otherwise it dies under mechanical regulation of growth. In this case, solving eq. \eqref{eq: elastic equation}, results in an exponential change of the metric with time: $g_{\mu\nu}(t)= e^{\gamma t} g_{\mu\nu}(0)$. 

\paragraph{Confined growth-} Finally, when a tissue proliferates against an external hard boundary, the pressure inside is expected eventually to be high enough  that the death rate cancels the active division rate at some finite pressure.  Thus we search for a homeostatic state in which $\pd_t Q^{\mu\nu} =0$. As the tissue is bounded by an external wall, a coordinate system in which $g_{\mu\nu} = \delta_{\mu\nu}$ may be chosen. As above, the existence of both internal and residual stresses inside the tissue is important and drives cells' death.

In the hard solid state, $\ptwo^0 <2$, the network tensor in the homeostatic state, $Q_h^{\mu\nu}$, exhibits a transition occurring at $\xi^*= \frac{(1+2H)(4-{\ptwo^0}^2)}{4\ptwo^0}$
\begin{align}\label{eq: homeostatic}
	\left. \begin{array}{cc}
		q_1=q_2 =\frac{\sqrt{\xi^2+(1+2H)^2(1+4k_P)(1+k_P {\ptwo^0}^2)}-\xi}{(1+2H)(1+4k_P)}, & \xi<\xi^* \\
		q_{1,2}=\frac{\ptwo^0}{2} \pm \frac{1}{2}\sqrt{{\ptwo^0}^2-4 + \frac{8\xi\left(\sqrt{\xi^2+\left(1+2H\right)^2}-\xi\right)}{(1+2H)^2}}  &\xi>\xi^*
	\end{array} \right.
\end{align}
When $\ptwo^0 \geq 2$ (no residual stress), however, there is no transition, and  $Q_h^{\mu\nu}$  is given by the second line of eq. \eqref{eq: homeostatic}, for all $\xi$. In both cases, the stability of the symmetry-broken solution ($q_2 = \ptwo^0 -q_1$) is dynamically driven, since for large $\xi$ values, it is energetically less favorable than the isotropic solution.
In fig. \ref{fig:pressure_active}, we plot the homeostatic pressure, defined as $p_h =- g_{\mu\nu}\sigma_{el}^{\mu\nu}$ when $Q^{\mu\nu}=Q^{\mu\nu}_h $.
\begin{figure}[h!]
	\centering
	\includegraphics[width=0.9\columnwidth]{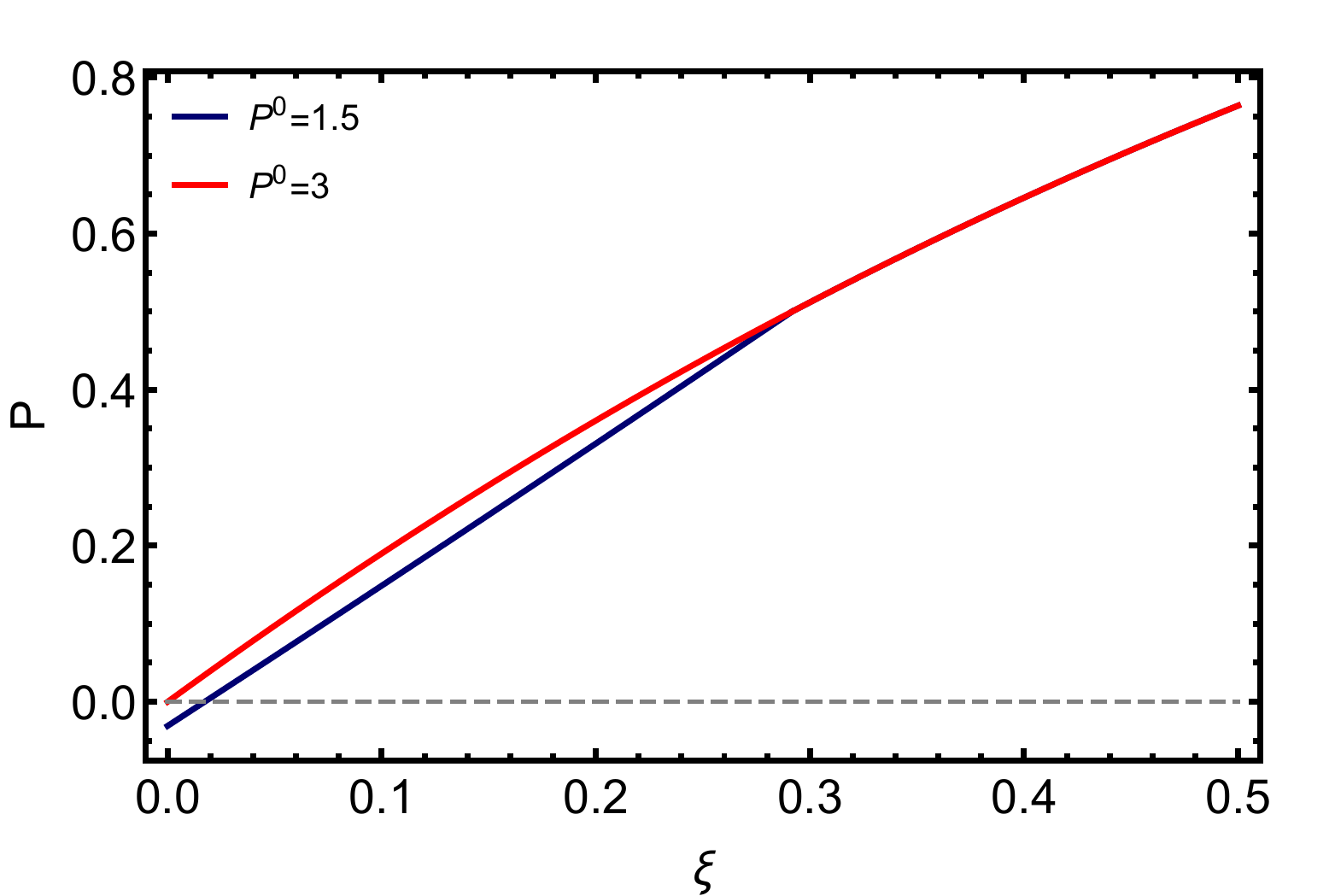}
	\caption{\label{fig:pressure_active} Homeostatic pressure of a tissue, as a function of the active division rate, $\xi$, for  $\ptwo^0=1.5 <2$ (blue) and $\ptwo^0 =3 >2$(red). When $\xi =0$, the pressure is negative for $\ptwo^0<2$ and vanishes for any $\ptwo^0 \geq 2$. Pressure for either $\ptwo^0>2$ or $\ptwo^0 <2$ whiel $\xi >\xi^*$ does not depend on $\ptwo^0$. Here $H=0$, $k_P=1$, $\xi^* \simeq 0.2917$. $p_h(\xi>\xi^*)$ is independent of $k_P$.   }
\end{figure}

Concluding, a wide and intricate range of complex behaviors that depend strongly on the existence of residual stresses at the cellular level is seen in this study. These exemplify the importance of microscopics on the macroscopic behavior of a tissue. A special focus should be given regarding the unique predictions of the homeostatic state.


Relaxing mean field, or adding  more complete Onsager relations are bound to shed new insights to the physics of living matter.  When uniformity is relaxed, the requirement that spatial derivatives vanish can almost always be achieved in many cases, recovering the usual defect pattern on sphere, for example.  Nevertheless, when considering evolving, non-flat geometries, one must include bending term, which are absent in this work.  

\paragraph{Acknowledgments}
This paper is dedicated to  the memory of a former teacher, Erez Barnoy, who recently passed away.


\bibliography{C:/Users/doron.grossman/Documents/PostDoc/Projects/Papers_and_Notes/bibs/active}

\clearpage

\onecolumngrid
\appendix
\section{Appendix A - Energy Functional}\label{appdx: Energy Functional}
Typically, the energy of the vertex model is written:
\begin{align}\label{appdxeq: vert}
	E & =  \sum_i  \hat{k}_A \left(A_i - \hat{A}^0_i\right)^2 + \tilde{k}_P \left(P_i -P^0_i\right)^2,
\end{align}
Focusing on the perimeter term:
\begin{align}\label{appdxeq: vert_ep}
	E_P & =  \sum_i  \tilde{k}_P \left(P_i -P^0_i\right) = \sum_i  \tilde{K}_P \left(\frac{P-P^0}{P^0}\right)^2  
\end{align}
where  we defined $\tilde{K}_P = \tilde{k}_P  {P^0}^2$ is the elastic modulus, and $\frac{P-P^0}{P^0}$ is the generalized strain along the perimeter. Since the strain is uniform along the perimeter, one can brake it onto smaller segments
\begin{align}\label{appdxeq: strain}
	 \left(\frac{P-P^0}{P^0}\right)^2  = \sum_{e} \left(\frac{\ell_e -\ell_e^0}{\ell_e^0}\right)^2
\end{align}
where $\ell_e$ is the actual length of an edge, and $\ell_e^0$ is a "reference" length, satisfying $\sum_{e }\ell_e^0 =P^0$. Thus we may write
\begin{align}\label{appdxeq: vert_ep_new}
	E_P & =  \sum_i  \sum_{e\in i} \tilde{K}_P \left(\frac{\ell_e-\ell_e^0}{\ell_e^0}\right)^2 ,
\end{align}
where the sum is taken over all edges associated with the $i^{th}$ cell. For small strains such that $\ell_e \sim \ell_e^0$ one may write approximate 
\begin{align}
	\frac{\ell_e - \ell_e^0}{\ell_e^0} \simeq \frac{\ell_e^2 - {\ell_e^0}^2}{2 {\ell_e^0}^2}.
\end{align}
Thus, finally 
\begin{align}\label{appdxeq: vert_ep_final}
	E_P & =  \sum_i  \sum_{e\in i} \tilde{K}_P \left(\frac{\ell_e-\ell_e^0}{\ell_e^0}\right)^2  \simeq  \sum_i  \sum_{e\in i} \tilde{K}_P \left(\frac{\ell_e^2-{\ell_e^0}^2}{4{\ell_e^0}^2}\right)^2 \\ \nonumber & \sum_i  \frac{\tilde{K}_P}{4{\ptwo^0}^2} \left(\frac{\ptwo_i-{\ptwo_i^0}^2}{{\ptwo_i^0}^2}\right)^2
\end{align}
where used the same uniform strain condition as before. This time under the constraint $\ptwo^0 = \sum_{e } {\ell_e^0}^2$. Redefining $k_P =  \frac{\tilde{K}_P}{4{\ptwo^0}^2} = \frac{\tilde{k}_P {{P^0}^2}}{4{\ptwo^0}^2}$, results in eq. \eqref{eq: vertex model 1}. Since we assume $\ell_0$ are given at this point, this should not pose any problem. However, eventually we relax this condition and only limit the sum. While limiting $\ptwo^0$ is not the same as limiting $P^0$, as both are distances after all.

\subsection{Appendix B - Derivatives in Coarse-Grained models}
Coarse-graining is the averaging of quantities over some scale $\ell$. This is done by taking a kernel function $\phi_\ell(x)=\phi_\ell(-x)$ so that $\int \phi_\ell(x) dx =1$ and $\phi_\ell(x)\xrightarrow{x\rightarrow \infty} 0$. Coarse-graining  a function $f(x)$ is given by the convulsion with the kernel-
\begin{align}
	f_\ell(x) =\int \phi_\ell(y)f(x-y) dy.
\end{align}
Assuming $\phi_\ell$ decays fast enough, relative to the scale of change of $f(x)$, one can write to second order in $\ell$-
\begin{align}
	f_\ell(x)  &=\int \phi_\ell(y)f(x-y) dy \simeq  \int \phi_\ell(y)\left[f(x) -y \pd_x f(x) + \frac{1}{2} y^2  \left(\pd_x f(x) \right)^2\right] dy =  f(x)+ C \ell^2  \left(\pd_x f(x) \right)^2,
\end{align}
where $C$ is some constant, assumed small.

The coarse-graining of the square of function gives a similar result-
\begin{align}
	(f^2)_\ell(x)  &=\int \phi_\ell(y)f^2(x-y) dy \simeq  \int \phi_\ell(y)\left(f(x) -y \pd_x f(x)\right)^2 dy =  f^2(x)+ C_2 \ell^2  \left(\pd_x f(x) \right)^2,
\end{align}
where $C_2$ is again some constant.
 
When coarse-graining the vertex model, we essentially start from a very singular cell distribution - the density, for example, is given by
$$\rho(x) = \sum \delta(x-x_i),$$ $x_i$ being the position of the $i^{th}$ cell.  Thus we assumes some scale over which a system realization is already coarse-grained. While in principle $\ell$ is an arbitrary scale, when it is large enough, the same expressions above results with a natural correlation scale  of the system (if exists) .
Therefore
\begin{align}
	\langle A^2 \rangle = \langle (A-A(x)+A(x))^2\rangle = \langle (\Delta A + A(x))^2\rangle =\langle \Delta A^2 \rangle + \langle (A(x))^2 \rangle =\langle \Delta A^2 \rangle +  (A(x))^2  +C_3 \ell^2 \left(\pd_x A(x)\right)^2,
\end{align}
where $C_3$ is some constant, and $\Delta$ .

\subsection{Appendix C - Full Coarse-Grained model}\label{appdx: Details}
The full Fokker - Planck equation is given by:
\begin{align}\label{eq: Fokker Planck Final_2}
	\pd_t \Psi =&   \frac{\pd^2}{\pd x^\mu \pd x^\nu} \left[D_{x}^{\mu\nu} \Psi\right]  \\ \nonumber
	& -\frac{\pd}{\pd A^0} \left[\tilde{K} \Psi \right] + \frac{\pd^2}{(\pd A^0)^2} \left[D_{\tilde{K}} \Psi\right] \\ \nonumber
	&  -\frac{\pd}{\pd \ptwo^0} \left[\tilde{\varepsilon} \Psi \right] + \frac{\pd^2}{(\pd \ptwo^0)^2} \left[D_{\tilde{\varepsilon}} \Psi\right] \\ \nonumber 
	& + \sum_i \left\{ -\frac{\pd}{\pd \Qtwo^{\mu\nu}}\left[\gamma_i \,\delta_i \Qtwo^{\mu\nu} \Psi\right] + \frac{\pd^2}{\pd \Qtwo^{\mu \nu}\pd \Qtwo^{\lambda \sigma}} \left[D_{i,\Qtwo}^{\mu\nu\lambda\sigma}\Psi\right]\right\}.
\end{align}
$D_x^{\mu\nu}$ is the total diffusion coefficient stemming from the different topological processes possible. The next lines describe the effect of noisy growth and topological transitions on the density function. Sum is taken over division ($d$), apoptosis ($a$), and $T_1$ transitions. $\tilde{K}$ is the average growth rate of a cell, $\tilde{\varepsilon}$ is the average elongation, and $\pd_t Q^{\mu\nu}$ is the average change in the network. $D_{\tilde{K}/\tilde{\varepsilon}/Q}$'s relate to noise. 

Beginning from eq. \eqref{eq: Fokker Planck Final_2} , we can compute the following dynamic equations for the variables $\alpha = \int A^0 \Psi(x,\chi) \dt \chi$ (where $\chi = \{A^0, \ptwo^0, \Qtwo\}$), $\lpow{2}{\tilde{\varpi}} = \frac{\int \ptwo_0 \Psi \dt \chi}{g_{\mu\nu}\Qtwo^{\mu\nu}\int \Psi \dt \chi}$

\begin{align}
	\label{eq: materials_eoms}
	\frac{\pd \alpha}{\pd t} &= \left(\gamma_a-\gamma_d\right) \alpha + \nabla_\mu\nabla_\nu \left[D_{x}^{\mu\nu} \alpha \right] + \tilde{K} \rho - \frac{\alpha}{2}g^{\mu\nu}\pd_t g_{\mu\nu} \\
	\frac{\pd \lpow{2}{\tilde{\varpi}}}{\dt t} &= \frac{\pd}{\pd t} \left(\frac{\lpow{2}{P}^0}{\lpow{2}{P}}\right) = \frac{\pd_t{\lpow{2}{P}}^0}{\lpow{2}{P}} - \frac{\lpow{2}{P}^0}{\lpow{2}{P}} \frac{\pd_t g_{\mu\nu}\Qtwo^{\mu\nu} + g_{\mu\nu} \pd_t \Qtwo^{\mu\nu} }{g_{\mu\nu} \Qtwo^{\mu\nu}} \\ \nonumber  &= \frac{1}{\lpow{2}{P}} \left( \tilde{\varepsilon} + \frac{\nabla_\mu \nabla_\nu \left(D^{\mu\nu}_x\rho \ptwo^0 \right)- \ptwo^0\nabla_\mu \nabla_\nu \left(D^{\mu\nu}_x\rho\right)}{\rho} \right) - \lpow{2}{\tilde{\varpi}} \frac{\pd_t g_{\mu\nu} \Qtwo^{\mu\nu} + g_{\mu\nu}\pd_t \Qtwo^{\mu\nu}}{g_{\mu\nu}\Qtwo^{\mu\nu}} \\
\end{align}
The metric $g$ is given by the instantaneous solution for the equation $$\nabla_\mu \sigma^{\mu\nu}_{el} =f_{ext}^\nu$$ where $\sigma^{\mu\nu}_{el} =\frac{\delta E}{\delta g_{\mu\nu}}$ is the stress tensor, and $f_{ext}^\nu$ is an external force acting on the tissue. Thus in each time step we solve for $g$ and allow $E$ to change via the change in $\Psi$.

Confluence of cells means $\rho = \frac{1}{\langle A \rangle}$. Using this relation and the time dependence of the  second moments-
\begin{align} 
	\pd_t \langle A_0^2 \rangle &= 2 \langle A_0 \rangle \tilde{K} + 2 D_{\tilde{K}} + \frac{1}{\rho} \left[ \nabla_\mu\nabla_\nu\left(D^{\mu\nu}_x\langle A_0^2 \rangle \rho\right)- \langle A_0^2 \rangle \nabla_\mu\nabla_\nu\left(D^{\mu\nu}_x \rho\right) \right]\\
	\pd_t \langle A^2 \rangle &=  \langle A \rangle^2  \left\{g^{\mu\nu}\pd_t g_{\mu\nu}- 2 \gamma - \frac{2}{\rho} \nabla_\mu\nabla_\nu\left[D^{\mu\nu}_x \rho\right] \right\} + \pd_t\left( \frac{\langle \Delta \rho ^2 \rangle}{\rho^4}\right) \\ 
	\pd_t \langle (\lpow{2}{P}^0)^2 \rangle &= 2 \langle  \lpow{2}{P}^0 \rangle \tilde{\varepsilon} + 2 D_{\tilde{\varepsilon}} + \frac{1}{\rho} \left[ \nabla_\mu\nabla_\nu\left(D^{\mu\nu}_x\langle (\lpow{2}{P}^0)^2 \rangle \rho\right)- \langle (\ptwo^0)^2 \rangle \nabla_\mu\nabla_\nu\left(D^{\mu\nu}_x \rho\right) \right]\\
	\pd_t \langle \lpow{2}{P}^2 \rangle &= 2\left( g_{\mu\nu}\pd_t g_{\alpha \beta} +2 g_{\mu\nu}g_{\alpha \beta} \right)\langle  \Qtwo^{\mu\nu} \pd_t \Qtwo^{\alpha \beta} \rangle+ 2 g_{\mu\nu}g_{\alpha \beta}D_Q^{\mu\nu\alpha\beta},
\end{align}
we can write the energy production rate. It is given fully by
\begin{align}\label{eq: Energy_prod}
	\frac{dE}{dt} = \int \left\{ \left( \frac{\pd \mathcal{E}}{\pd t}\right)_g +\left( \frac{\pd \mathcal{E}}{\pd t}\right)_{T_1} + \left( \frac{\pd \mathcal{E}}{\pd t}\right)_\gamma  +\left( \frac{\pd \mathcal{E}}{\pd t}\right)_{growth} + \left( \frac{\pd \mathcal{E}}{\pd t}\right)_{disorder} +\left( \frac{\pd \mathcal{E}}{\pd t}\right)_{diff} \right\}\rho \sqrt{g} \dt x
\end{align}
where
\begin{align}
	\label{eq:energy_prod_elastic} &\left( \frac{\pd \mathcal{E}}{\pd t}\right)_g  & = &\,\left[K_A \left(1-\alpha\right) g^{\mu\nu} + 2 K_P \left( 1- \varpitwo\right) \frac{Q^{\mu\nu}}{\ptwo}\right] \pd_t g_{\mu\nu}  \\
	\label{eq:energy_prod_T1} &\left( \frac{\pd \mathcal{E}}{\pd t}\right)_{T_1}  & = &\, 2 K_P \left(1-\varpitwo\right) \frac{g_{\mu\nu}}{\ptwo} ~\left( \pd_t Q^{\mu\nu}\right)_{T_1} \\
	\label{eq:energy_prod_prilferatin} &\left( \frac{\pd \mathcal{E}}{\pd t}\right)_{\gamma_{d/a}}  &= &\, \left[-K_A \left(1-\alpha^2\right) + K_P \left(1-\varpitwo\right)^2 \right]\gamma_{d/a} +2 K_P \left(1-\varpitwo\right) \frac{g_{\mu\nu}}{\ptwo} \left(\pd_t Q^{\mu\nu}\right)_{\gamma_{d/a}}\\
	\label{eq:energy_growth} &\left( \frac{\pd \mathcal{E}}{\pd t}\right)_{growth}  & = &\, -2 K_A \left(1-\alpha\right) \rho \tilde{K}  - 2 K_P \left(1-\varpitwo \right)\frac{\varepsilon}{\ptwo} +  2 K_A D_{\tilde{K}} \rho^2 + 2 K_P \frac{D_{\tilde{\varepsilon}}}{\ptwo^2}  \\
	\label{eq:energy_prod_disorder} &\left( \frac{\pd \mathcal{E}}{\pd t}\right)_{disorder}  & = &\, - K_A \rho^{2} \pd_t \left(\frac{\langle \Delta \rho^2 \rangle}{\rho^4}\right) + 2 K_P  \frac{\langle \Delta Q^{\mu\nu} \Delta Q^{\alpha \beta} \rangle}{\ptwo^2} g_{\mu\nu} \pd_t g_{\alpha \beta} + 2K_p \frac{g_{\mu\nu} g_{\alpha \beta}}{\ptwo^2} \langle \Delta Q^{\mu\nu} \pd_t Q^{\alpha \beta} \rangle\\ \nonumber
	&&&\, +\left[ K_A \rho^2 \left( \langle \Delta \rho^2 \rangle + \langle \Delta A_0^2 \rangle \right) +  K_P \frac{1}{\ptwo^2}\left( \langle \Delta \ptwo^2 \rangle + \langle \Delta \ptwo_0^2 \rangle \right) \right] \left(\gamma_d-\gamma_a\right) \\
	\label{eq:energy_prod_diffusion} &\left( \frac{\pd \mathcal{E}}{\pd t}\right)_{diff}  &= &\, \left(-K_A \left(1-\alpha^2\right) + K_P \left(1-\varpitwo\right)^2\right) \frac{1}{\rho}\nabla_\mu \nabla_\nu \left(D^{\mu\nu} \rho \right) \\ 
	\nonumber &&&\, + K_A \left[\rho \mathcal{D} \left(\langle A_0^2 \rangle \right) - 2 \mathcal{D}\left(\langle A_0 \rangle \right) \right]
	+ K_P \left[\frac{1}{\rho \ptwo^2} \mathcal{D}\left( \langle P_0^2 \rangle  \right) - 2 \frac{1}{\rho \ptwo}\mathcal{D}\left(\langle P_0 \rangle \right) \right]
\end{align}
Here $K_A = k_A g Q$, $K_P = k_P \left(g_{\mu\nu}Q^{\mu\nu}\right)^2$, and $\gamma_i = - \frac{1}{2}Q^{-1}_{\mu\nu}\pd_t Q^{\mu\nu}_i$. Additionally defined the "covariant" diffusion functional, of a scalar $f$ is 
\begin{align}
	\mathcal{D}\left(f\right) =\nabla_\mu\nabla_\nu\left(D^{\mu\nu}\rho f\right) - f \nabla_\mu\nabla_\nu\left(D^{\mu\nu}\rho \right),
\end{align}
which describes how a value $f$, carried by the cell density, diffuses regardless of the effect of diffusion on the density $\rho$. Indeed, when $f$ is position independent, this terms is zero.

Equation \eqref{eq: Energy_prod} can be written in a simple form 
\begin{align}\label{eq: Enery_prod_simp_2}
	\frac{dE}{dt} = \int  \left\{\gamma \mu - \sigma_{el}^{\mu\nu}\pd_t g_{\mu\nu} - \sigma_A \tilde{K} - \sigma_P \tilde{\varepsilon} + \eta_{\beta \gamma} \pd_t \Qtwo^{\mu\nu}+ \eta_D \right\} \rho\sqrt{g} \dt x
\end{align}
where $\mu$ can be interpreted  as a "chemical potential", $\sigma_{el}^{\mu\nu}$ is the elastic stress, $\sigma_A$ is an active "pressure", $\sigma_P$ is an active "line tension", $\eta_{\beta \gamma}$ is the flow - stress, and $\eta_D$ is the energy dissipation due to diffusion, and noise. They are given by (omitting  "disorder" terms for simplicity)
\begin{align}
	\mu &= -K_A \left(1- \alpha^2 \right) + K_P \left(1-\lpow{2}{\tilde{\varpi}}\right)^2 + -2 K_P \left(1-\varpitwo\right) \frac{g_{\mu\nu}}{\ptwo} \frac{\left(\delta \Qtwo^{\mu\nu}\right)}{Q^{-1}_{\alpha \beta}\pd_t Q^{\alpha\beta}} \\ 
	\sigma_{el}^{\mu\nu} &= -K_A \left(1 - \alpha \right)g^{\mu\nu} -  2K_P\left(1-\lpow{2}{\tilde{\varpi}}\right)\frac{\Qtwo^{\mu\nu}}{\ptwo} \\
	\sigma_A &= -2 K_A \rho \left(1-\alpha\right)\\ 
	\sigma_P &= -2 K_P \frac{1}{\lpow{2}{P}}\left(1-\lpow{2}{\tilde{\varpi}}\right)\\
	\eta_{\mu \nu} &= 2K_P \left(1-\varpitwo\right)\frac{g_{\mu\nu}}{\ptwo} \\ 
	\eta_D & = \left(-K_A \left(1-\alpha^2\right) + K_P \left(1-\varpitwo\right)^2\right) \frac{1}{\rho}\nabla_\mu \nabla_\nu \left(D^{\mu\nu} \rho \right) \\ 
	\nonumber &\, + K_A \left[\rho \mathcal{D} \left(\langle A_0^2 \rangle \right) - 2 \mathcal{D}\left(\langle A_0 \rangle \right) \right]
	+ K_P \left[\frac{1}{\rho \ptwo^2} \mathcal{D}\left( \langle P_0^2 \rangle  \right) - 2 \frac{1}{\rho \ptwo}\mathcal{D}\left(\langle P_0 \rangle \right) \right]
\end{align}

Using an Onsager approach, we may now relate the unknown fluxes ($\pd_t Q^{\mu\nu}, \tilde{K}, \tilde{\varepsilon}$) to the potentials $\mu,\sigma_{el},\eta_{\mu\nu}$ which have an explicit expression using our variables.  Such models in general could be very complex.


\subsection{Appendix D -Choice of $\Gamma^{\mu\nu\alpha\beta}$}
At this point we are missing explicit relations between the rates and and the stress (or other expressions). One way to do so be to apply a minimal Onsager appraoch so that the rate of change  is linear in its conjugate, with the least couplings. Thus, a naive guess would be
\begin{align}\label{eq:rates_onsager}
	\left(\pd_t \Qtwo^{\mu\nu}\right)_i = \Gamma^{\mu\nu\alpha\beta}_i \frac{\delta E}{\delta Q^{\alpha\beta}}
\end{align}
where the second term appear since $\rho$ is (non-trivially) related to $\Qtwo^{\mu\nu}$. $\Gamma_i^{ \mu\nu\alpha \beta}$ is a symmetric Onsager tensor relating the geometry to changes in network.

It is immediately clear that $\Gamma^{\mu\nu\alpha\beta}$ is symmetric to exchange of $\mu \leftrightarrow \nu$ and $\alpha \leftrightarrow \beta$. From the contribution to the energy generation rate, it is also clear that it must be symmetric to $(\mu,\nu) \leftrightarrow (\alpha,\beta)$. 

The most general expression satisfying the these symmetries is 

\begin{align}
	\Gamma^{\mu\nu\alpha\beta} & =\frac{H_1}{2}\left(\Qtwo^{\mu\alpha}\Qtwo^{\nu\beta} + \Qtwo^{\nu\alpha}\Qtwo^{\mu\beta}\right)+ H_2 \Qtwo^{\alpha \beta}\Qtwo^{\mu\nu} \\ \nonumber
	&+\frac{H_3}{2}\left((\Qtwo^{-1})^{\mu\alpha}(\Qtwo^{-1})^{\nu\beta} + (\Qtwo^{-1})^{\nu\alpha}(\Qtwo^{-1})^{\mu\beta}\right)+ H_4 (\Qtwo^{-1})^{\alpha \beta}(\Qtwo^{-1})^{\mu\nu} \\ \nonumber
	&+\frac{H_4}{4}\left(\Qtwo^{\mu\alpha}(\Qtwo^{-1})^{\nu\beta} +(\Qtwo^{-1})^{\mu\alpha}\Qtwo^{\nu\beta} + \Qtwo^{\nu\alpha}(\Qtwo^{-1})^{\mu\beta}+(\Qtwo^{-1})^{\nu\alpha}\Qtwo^{\mu\beta}\right)+ \frac{H_5}{2}\left( \Qtwo^{\alpha\beta}(\Qtwo^{-1})^{\mu\nu}+(\Qtwo^{-1})^{\alpha \beta} \Qtwo^{\mu\nu} \right) \\ \nonumber
	&+\frac{G_1}{2}\left(g^{\mu\alpha}g^{\nu\beta} + g^{\nu\alpha}g^{\mu\beta}\right)+ G_2 g^{\alpha \beta}g^{\mu\nu} \\ \nonumber
	&+\frac{G_3}{4}\left(g^{\mu\alpha}\Qtwo^{\nu\beta}+\Qtwo^{\mu\alpha}g^{\nu\beta} + g^{\nu\alpha}\Qtwo^{\mu\beta}+\Qtwo^{\nu\alpha}g^{\mu\beta}\right)+ \frac{G_4}{2}\left( g^{\alpha \beta}\Qtwo^{\mu\nu}+\Qtwo^{\alpha \beta}g^{\mu\nu}\right) \\ \nonumber
	&+\frac{G_5}{4}\left(g^{\mu\alpha}(\Qtwo^{-1})^{\nu\beta}+(\Qtwo^{-1})^{\mu\alpha}g^{\nu\beta} + g^{\nu\alpha}(\Qtwo^{-1})^{\mu\beta}+(\Qtwo^{-1})^{\nu\alpha}g^{\mu\beta}\right)+ \frac{G_6}{2}\left( g^{\alpha \beta}(\Qtwo^{-1})^{\mu\nu}+(\Qtwo^{-1})^{\alpha \beta}g^{\mu\nu}\right) \\ \nonumber
\end{align}

However, terms including the $Q^{-1}$, have negligible contributions for large $Q$'s and are thus unlikely. In any case, just as with terms involving $g^{\mu\nu}$,  these  expressions cannot contribute to a $T_1$ transition as they they contribute either isotropically ($g^{\mu\nu}$) or inversely ($Q^{-1}$, making shorter dimensions even shorter ). We thus adopt a minimal scheme-
\begin{align}\label{eq: H_tens_coeff}
	\Gamma^{\mu\nu\alpha\beta} =H_1\left( \frac{1}{2}\left(\Qtwo^{\mu\alpha}\Qtwo^{\nu\beta} + \Qtwo^{\nu\alpha}\Qtwo^{\mu\beta}\right)+ H \Qtwo^{\alpha \beta}\Qtwo^{\mu\nu}\right).
\end{align} 
where $H_2 = H_1 H$
This suggest that $\pd_t \Qtwo ^{\mu\nu} \propto H_1( \Qtwo^{\mu}_\alpha \Qtwo^{\alpha \nu} + H \Qtwo^{\mu\nu}\ptwo)$, thus, a suitable choice of the coefficient $H$  can account for different transitions. Without loss of generality, one can set $H_1=1$. The choice
\begin{align} \label{eq: T_1_H_coeff}
	H=-1/2
\end{align} results, at the locally flat, Q - diagonal frame, where $Q = Diag[q_1, q_2]$,  with $\pd_t Q \propto \frac{1}{q_1+q_2}Diag[q_1(q_1- q_2),-q_2(q_1- q_2)]$ and $\pd_t E \propto (q_1-q_2)^2$  is expected for a $T_1$ transition.

Since $\rho$ and $\Qtwo$ are related, our knowledge of the dynamics of $\rho$ can be used determine $H_2$ by the flow and fluctuation terms to those derived from $\pd_t \Qtwo^{\mu\nu}$.

For $H=0$, at the locally flat, $\Qtwo$ - diagonal  frame $\left(\delta \Qtwo^{\mu\nu}\right)_\gamma \propto Diag[q_1^2, q_2^2]/(q_1+q_2)$. Note that the choice of $\left(\delta \Qtwo^{\mu\nu}\right)_\gamma $ results in a negative contribution to the energy whenever $\gamma>0$ and $\varpitwo<1$ (perimeter too long), or $\gamma <0$ and $\varpitwo>1$ (too short). While for the cases of $\gamma>0$ and $\varpitwo>1$, or $\gamma <0$ and $\varpitwo<1$, this contribution is negative but small. The overall contribution must be negative of course. 
This is in qualitative agreement with a realistic case (since division only shortens, and appoptosis lengthens), but quantitatively it differs. Consider the case of division when the cells are too short, realistically we will only change the short dimension of the cell so as to not loose too much length, while in this modelling shortening happens in both dimensions.

\subsection{Appendix E - External Force}

We now turn to solve the case of finite sudden strain. And we will work within the mean field approximation (where our dynamical fields, $g_{\mu\nu}$ and $Q^{\mu\nu}$ do not vary through space). We thus consider a square tissue, with some initial $Q^{\mu\nu}$, located such that one edge is given by the coordinate $y=0$, the other by $y= Y$, and similarly there's an edge at $x=0$ and $x=X$.  At time zero we set the tissue's configuration $\vec{f}(x,y)$ so that $\left| \vec{f}(x=X) -\vec{f}(x=0)\right| = L$. We can incorporate this constraint to an effective energy  using a Lagrange multiplier
\begin{align}\label{eq:sudden strain}
	E_{ext} & = \int  \lambda \left(\left| \vec{f}(x=X) -\vec{f}(x=0)\right|-L\right) \dt y = \int \dt y \lambda \left(\left| \int\limits_0^X \pd_x\vec{f} \dt x\right|-L\right) \dt y \\ \nonumber
	& =  \int  \lambda \left(\sqrt{\int \int  \vec{f}'(x) \cdot \vec{f}'(x')  \dt x' \dt x }-L\right) \dt y .
\end{align}
Within the mean field approach $\vec{f}'(x) = \pd_x \vec{f}](x) = const.$ So that $\vec{f}'(x) \cdot \vec{f}'(x')= g_{11}$ we can write
\begin{align}
	E_{ext} & =\int  \lambda \left(\sqrt{X^2 g_{11} }-L\right) \dt y = \int  \lambda \left(X \sqrt{ g_{11} }-X \sqrt{G}\right) \dt y=  \int   {\lambda} X \left(\sqrt{ g_{11} }-\sqrt{G}\right) \dt y \\\nonumber
	&= \int  {\lambda} \left(\sqrt{ g_{11} }-\sqrt{G}\right) \dt x \dt y  
\end{align}
where $G$ is the length scale of the stretched tissue. 

Taking the variation with respect to $g$ an $\lambda$ we derive the following equations:
\begin{align}\label{eq:sud strain EOM}
	\sqrt{g_{11}} = \sqrt{G} \\ \nonumber
	\sigma^{\mu\nu} = \sigma_{el}^{\mu\nu} + \frac{\lambda}{2 \sqrt{g_{11}}} \left(\begin{array}{cc}
		1 & 0\\
		0 & 0 
	\end{array}\right)
\end{align}
where $\sigma_{el}^{\mu\nu}$ is the elastic stress given above.
In principle, in order to find the metric $g_{\mu\nu}$ minimizing the elastic energy we need to find the stress so that on the boundaries $x=0$ and $x=X$, the stress balances the force.  In the mean field approximation the conditions at the boundary impose $\sigma^{\mu\nu}= 0$.

We start by assuming  $Q=  diag(q_{1},q_{2})$ and $g= diag(g_{11}, g_{22})$, are aligned along the principal stretching direction. Finding $g_{22}$ and $\lambda$ by solving eq. \eqref{eq:sud strain EOM} for given $q_1, q_2, G$, we can explicitly solve the problem. By defining strain  with respect to the free (non stretched) tissue- $\Delta = g_{11} - g_{11}^{free}$. We find that compatible systems ($P^0 >2$) exhibit a bi-stability corresponding to exchange of $g_{11}^{free}$ and $g_{22}^{free}$, i.e - for $\Delta^*$ = $g_{22}^{free}-g_{11}^{free}$. In figure \ref{fig: Energy forced} we plot the elastic energy as a function of $\Delta$, for three cases - compatible case $P^0 >2$ (green), marginal case $P^0=2$ (orange), and incompatible case $P^0<2$. The fact that $E(\Delta =0, P^0 <2) \neq 0 $ is a clear sign of residual stresses.

\begin{figure}[!h]
	\centering
	\begin{tabular}{cc}
		\begin{subfigure}{0.5\textwidth}
			\centering
			\includegraphics[width=0.95\linewidth]{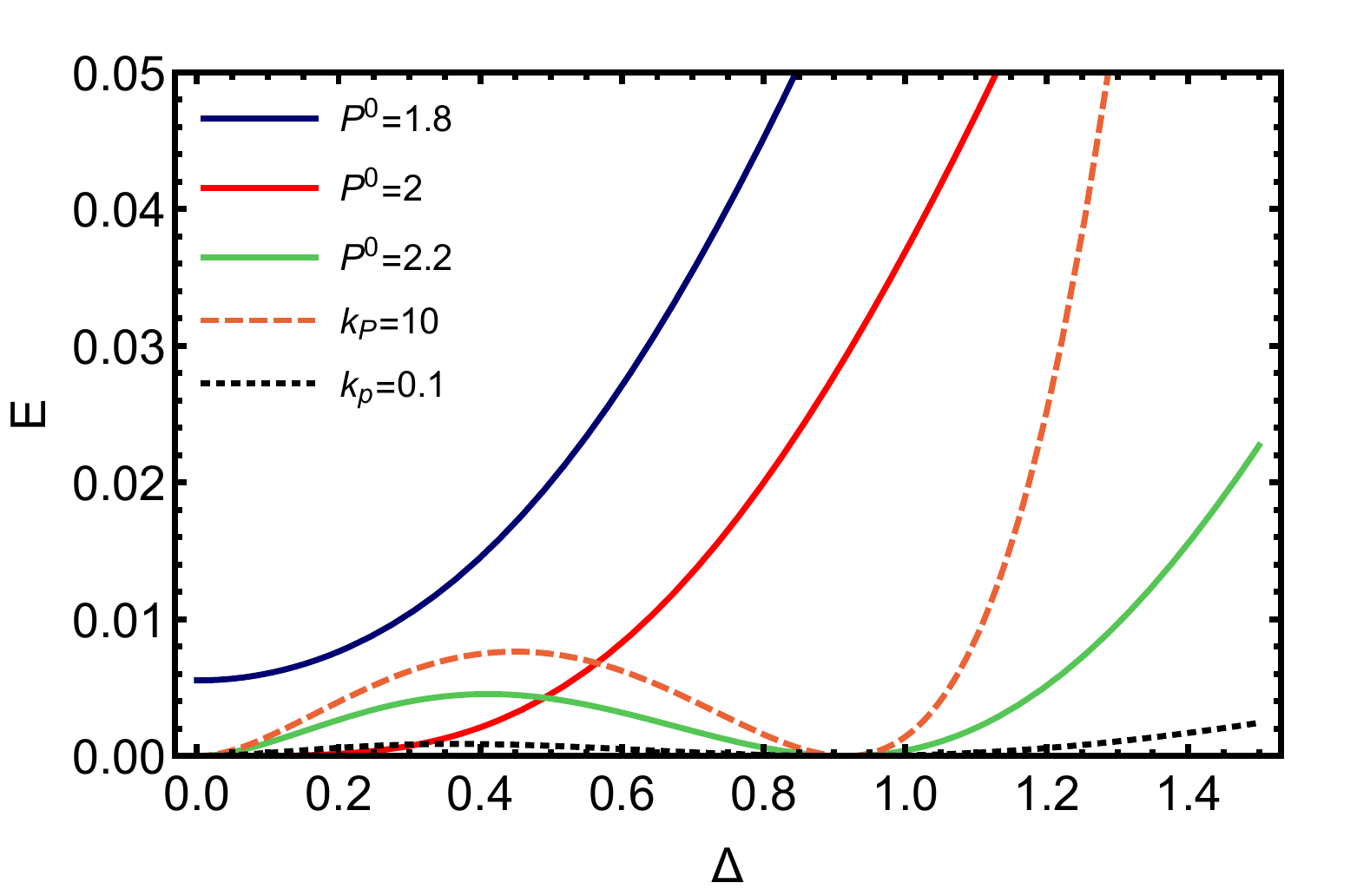}
			\caption{\label{fig: Energy forced}}
		\end{subfigure} &
		\begin{subfigure}{0.5\textwidth}
			\centering
			\includegraphics[width=0.95\linewidth]{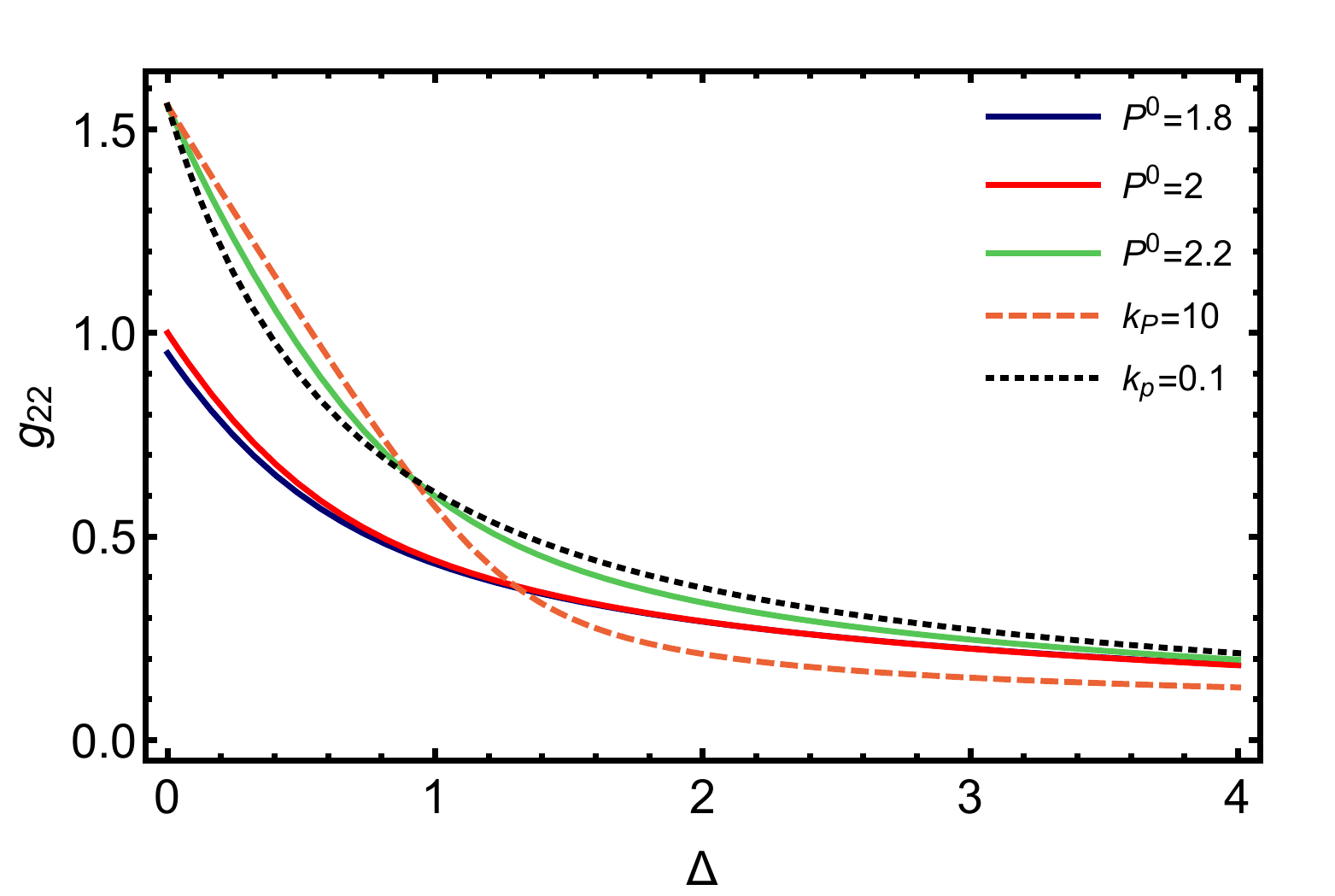}
			\caption{\label{fig: g22}}
		\end{subfigure} \\
		\begin{subfigure}{0.5\textwidth}
			\centering
			\includegraphics[width=0.95\linewidth]{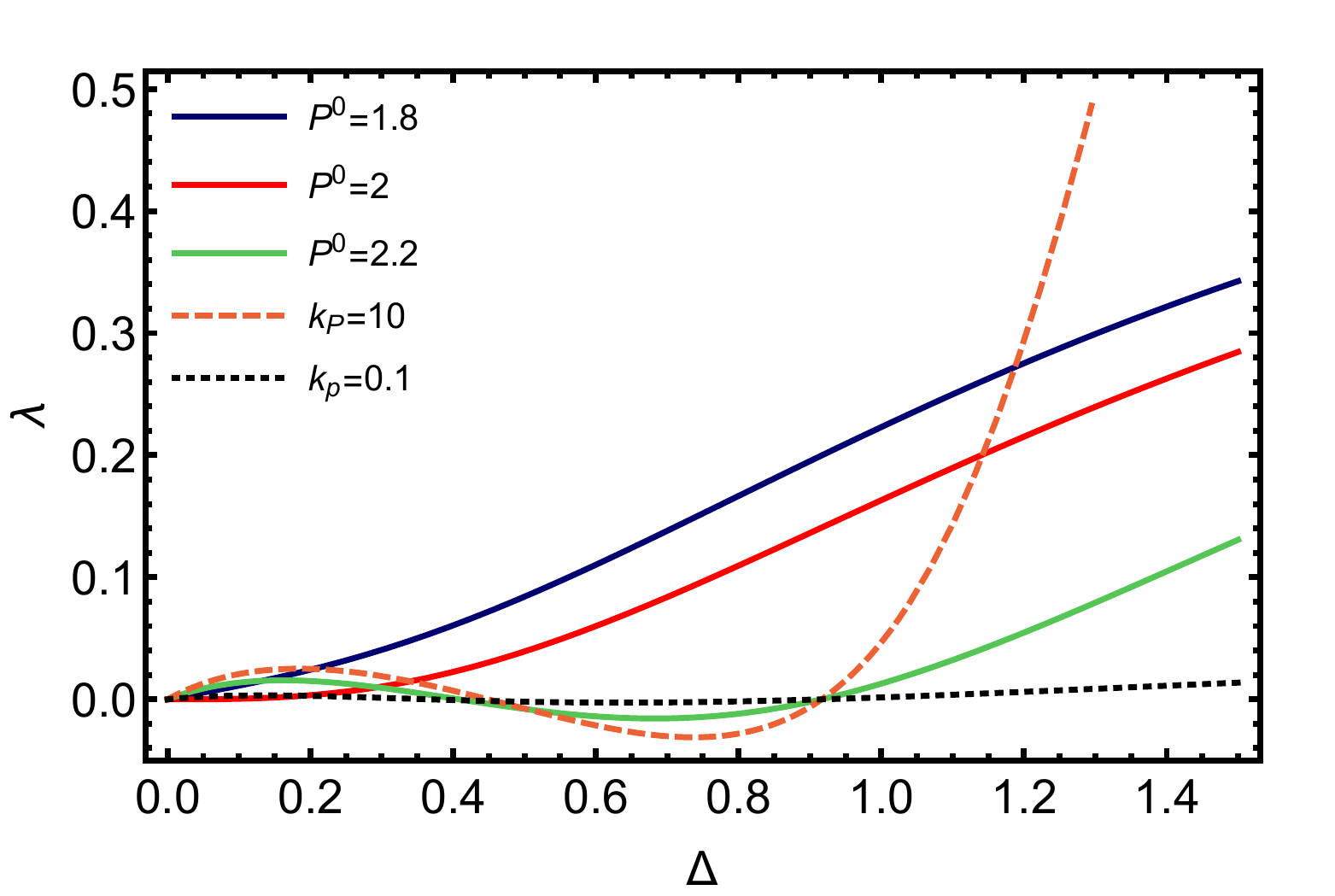}
			\caption{\label{fig: force}}
		\end{subfigure} &
		\begin{subfigure}{0.5\textwidth}
			\centering
			\includegraphics[width=0.95\linewidth]{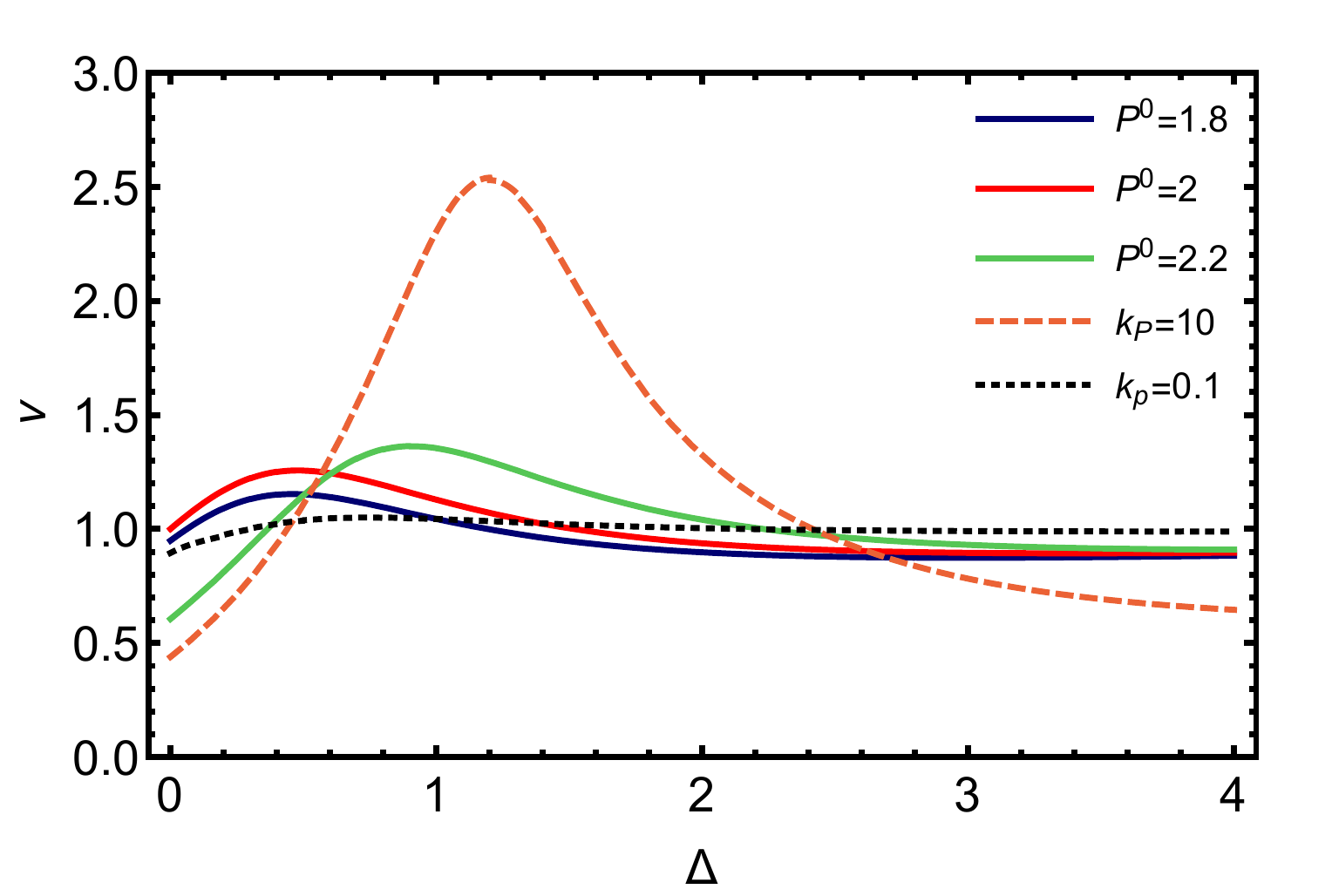}
			\caption{\label{fig: Poisson's Ratio}}
		\end{subfigure} \\
	\end{tabular}
	\caption{Tissue energy (a), $g_{22}$ (b), force (c), and Poisson's ratio ($\nu$, d), for the compatible case (green line), marginal case (red), and incompatible case(blue), as indicated in the legend, all with $k_p =1$. Additionally, different values of $k_p$ are plotted (dashed orange, and dotted black lines) for $P^0=2.2$ (compatible). As a function of  $\Delta = g_{11}-g_{11}^{free}$\label{fig:forced_solid}. Note that for the solid green, dashed red, and dotted graphs (all with $\ptwo_0>2$), there are two energetically favored states - with $\Delta =0 $, and $\Delta=\Delta^*$, which in this case $\delta* \simeq 0.9165$}
\end{figure}

Poisson's ration is typically defined as the derivative of the transverse strain vs the longitudinal strain $ \nu = - \frac{d\epsilon_y}{d\epsilon_x}$, near mechanical equilibrium. In our notation this translates into 
\begin{align}\label{eq: Poissons_ratio}
	\nu &= -\frac{d \log g_{22}}{d \log \Delta}
\end{align} 
$\Delta = g_{11}-g_{11}^{free}$ the tissue elongation.
Calculating this value for arbitrarily large differences of $g_{11}$, yields figure \ref{fig: Poisson's Ratio}. The most interesting feature of it is that far from $\Delta=0$, Poisson's ratio may be arbitrarily large on some finite region of $\Delta$. Not surprisingly, for $\Delta \rightarrow \infty$, $\nu \rightarrow 1$.

%
%
\subsection{Appendix F - Relaxation of non-active tissue via $T_1$ transition}
When only $T_1$ transitions are allowed, $H= -\frac{1}{2}$, we look for the stationary solutions of
\begin{align}
	\dot{q}_i = &- \frac{2}{\sqrt{q_1 q_2}} k_p \left(q_1+q_2-P^0\right)\left[q_i -\frac{1}{2} \left(q_1 + q_2\right) \right] q_i.
\end{align}
Those are characterized by the initial area  $A(0)= \sqrt{q_1(0) q_2(0)}$, rather than $\ptwo^0$. 
\begin{align}
	\left. \begin{array}{cc}
		q_1=q_2 =A(0), & A(0)> \frac{P^0}{2} \\
		q_1 = \frac{{A(0)}^2}{q_2} = \frac{1}{2}\left(P^0 \pm \sqrt{{P^0}^2-4 {A(0)}^2}\right), & A(0) \leq \frac{P^0}{2}
	\end{array} \right.
\end{align}

\end{document}